# Opportunities in deep learning methods development for computational biology

Alex J. Lee[1], Reza Abbasi-Asl[1,#]

[1]University of California, San Francisco

[#]Corresponding author: Reza Abbasi-Asl (Email: Reza.AbbasiAsl@ucsf.edu)

## Abstract

Advances in molecular technologies underlie an enormous growth in the size of data sets pertaining to biology and biomedicine. These advances parallel those in the deep learning subfield of machine learning. Components in the differentiable programming toolbox that makes deep learning possible are allowing computer scientists to address an increasingly large array of problems with flexible and effective tools. However many of these tools have not fully proliferated into the computational biology and bioinformatics fields. In this perspective we survey some of these advances and highlight exemplary examples of their utilization in the biosciences, with the goal of increasing awareness among practitioners of emerging opportunities to blend expert knowledge with newly emerging deep learning architectural tools.

## Introduction

The profusion of molecular and functional profiling techniques for biological systems is producing an enormous amount of data. Correspondingly, there is a more urgent need to leverage new data analysis tools to turn this enormous amount of data into insights and theories. Large scale and distributed projects such as the Human Cell Atlas[1] and BICCN/BICAN initiative[2] are generating enormous amounts of data across a variety of modalities such as single-cell RNA sequencing (scRNA-seq), spatially resolved transcriptomics (ST), and electrophysiology, allowing researchers to study at multiple levels (molecular, anatomical, functional) the complexity of organ systems such as the brain.

These advances parallel a revolution in computational analysis methods encompassing both statistics and machine learning (ML), and in particular the deep learning subfield of machine learning (DL) has paralleled advances in experimental methods development. Increasing recognition of the capacity of these methods to improve various aspects of the scientific discovery process as a whole[3,4] hints at the tremendous opportunities for data science, which we will loosely refer to as the combination of the areas of statistics, ML, and DL, to advance research in the biosciences.

The view that application of complex computational models to increasingly large and complex datasets will lead to meaningful understanding of complex systems may be overly optimistic[5]. Insofar as the success of ML and DL tools has largely been supervised learning problems with very large amounts of data, it is unsurprising that AlphaFold2[6], which was made possible by large amounts of public data[7], remains one of the most salient example of DL applied to the biosciences.

However, what is unprecedented is the ease with which researchers can access a large and powerful toolbox of differentiable deep learning programming software such as PyTorch[8] and Jax[cite], as well as the increasing flexibility of this toolbox to address a wide variety of modeling problems. Kaznatcheev and Kording[9] describe the "evolutionary" development of increasingly flexible and powerful components of the DL toolbox. In particular, they describe how the tooling of DL, meaning the collection of optimization algorithms, objective functions, and [something else], can be interchanged amongst each other without failure. Said in another way, recent innovations in DL make it easier to apply DL to a variety of different tasks with greater effectiveness.

In this perspective, we will discuss new opportunities to use the toolbox of DL to better incorporate inductive bias and prior knowledge into models, assuming a rudimentary understanding of ML and DL. The goal of this work is to highlight recent innovations in fundamental DL technologies and their potential to improve and assist biological discovery research.

## Recent expansion in the building blocks of deep learning architectures and training schemes enables new directions in computational biology

The development of differentiable differentiation programming frameworks have helped establish an array of general architectural building blocks and training schemes for DL models. These components have allowed practitioners to augment the traditional pattern of composing matrix operations with nonlinearities with more efficient and flexible subcomponents. Often these building blocks can also incorporate useful inductive biases, such as in graph neural networks (GNNs), where entity dependencies (for example, between atoms in a molecule) can be encoded via a combination of learned and fixed user-specified mechanisms. In the simple example of molecular property prediction, this might take the form of a neural network where at each layer a weighted sum is taken of features from different atoms that have bonds between them.

The concept of an inductive bias can be variably defined, but can be thought of as a practitioner choice that increases the likelihood of a model to extract a specific pattern of correlations from a dataset. In the molecular property prediction example, a non-GNN model might take as inputs the number of carbons, oxygens, and nitrogens in a given molecule and predict its solubility. This model does not directly incorporate notions of different connectivity, and would be unable to distinguish, for example, 1-proponal vs isopropyl alcohol, which have the same molecular formula of $C_3H_8O$. A model such as a GNN might be said to incorporate a useful inductive bias by allowing features of specific atoms to interact based on whether bonds are between them, mimicking the actual structure of the molecule.

The idea of extracting specific patterns of correlations that may be partially determined by the practitioner is a useful concept from which to describe a variety of recent innovations in DL architectural components and training schemes. In the next sections we will discuss two particularly important innovations, self- and semi-supervised training, and the attention mechanism, in terms of their ability to enable the incorporation of novel inductive biases in the modeling process, and their potential to improve computational biology modeling efforts.

## Incorporating assumptions on the semantics of latent representations using self-supervised learning

Data that are labeled at the scale which would be useful for supervised DL is relatively scarce. In this context, it is useful to consider other regimes for training DL models, such as self-supervised learning (SSL). Although SSL is not new, in the presence of large (unlabeled) datasets, it has significantly boosted many modern DL efforts[10]. In this section we will describe how SSL is a general framework for constructing optimization problems that can yield useful latent representations.

One of the easiest to understand methods in SSL is contrastive SSL, sometimes historically referred to as deep metric learning. Triplet loss is perhaps the most intuitive example of contrastive SSL: given a given reference data instance, a network is trained to compute a latent embedding which is optimized such that a positive example (for example a data point with the same class label) is close in latent space and a negative example (different class label) is far. This is in contrast to the traditional paradigm of supervising a network to predict a class label, or an unsupervised approach exemplified by methods such as autoencoders which are optimized for compression and reconstruction. If applied to single cell sequencing data, a positive example might be a cell with the same disease label as the reference cell and a negative example a cell with a different disease label. Another example of SSL is masked image modeling, where a network is trained to predict unseen image patches from observed ones.

One way to understand these methods is in seeing them as methods for grouping semantically similar data points together in the latent representation space of the model. Triplet loss can be seen as a way of preserving class label discreteness in latent space, and a masked image modeling paradigm can be interpreted as a method for discovering high-level representations that are invariant to patch loss[11]. Using the previous analogy of capturing correlations, SSL techniques can be interpreted as preserving certain correlations depending on their relevance to the specific objective function.

CEBRA[12], a recent work, showcases some possible advantages of a contrastive learning framework. Specifically, CEBRA can be used to discover latent variables for behavior (physical position of an animal in space) and neural activity (calcium imaging) simultaneously by using multiple variational autoencoders (VAEs), one for each modality. These VAEs are trained such that latent features corresponding to time points that are close together are close together in latent space, or optionally so that latent features corresponding to physical distance are likewise close together when the physical positions are similar. These latent features are shown to be useful for decoding behavioral dynamics as well as multimodal data integration.

An important note is that the choice of which factor is used to supervise the training process (either position, time, or a behavioral categorical label for example) amounts to an assumption that the latent variables from the network are conditionally independent given information on the supervision factor. By offering practitioners a powerful means by which to structure the model optimization process, this framework can easily be extended to a variety of applications. For example, CEBRA showcases how a user that is interested in animal-invariant features in the data can train the network across animals and hypothesize that such a scheme forces the network to identify features that are consistent amongst subjects.

Another illustrative example of SSL comes from the NCEM method[13]. This method uses a GNN that predicts a cell's observed gene expression vector (measured for example by MERFISH) from its cell type label and features from surrounding cells in the same neighborhood. In this way it is similar to the masked imaging modeling paradigm, but with unseen gene expression instead of unseen image patches. NCEM is an excellent example of utilizing the machinery of SSL to answer high-level biological questions. Specifically, the problem of understanding how far away cells need to be in order to be predictive of gene expression is formulated as a masked prediction problem. The authors explore the impact of varying neighborhood size (the radius at which a cell is considered within the neighborhood) and quantifying its effect on predictive accuracy. Using this framework to quantify cell-cell interaction, they also uncover a dependency of immune cell abundance to proximity to the tumor-immune boundary. The authors also parameterize a spatial sender-receiver interaction matrix that is optimized in the same process that provides a mechanism to inspect cell-cell interaction. Thus in NCEM the formulation of a general prediction problem on cellular neighborhoods is used as a way to inspect the relative contributions of spatial distance on cell-cell interactions.

These methods will continue to be broadly useful for injecting human biases into optimization problems and create generally useful representations. The spatial context of ST lends itself to a variety of relatively simple priors that can be implemented via SSL. For example, NCEM implicitly defines a prior that cellular communication ought to be on a relatively short distance scale. In an alternative construction, a model could be trained to create latent features for a cellular neighborhood in ST data and a contrastive loss could be used to optimize latent features for neighborhoods nearby in space to be close in latent space. These techniques provide biologists a useful toolkit for extracting meaningful representations from unlabeled data and operating on them. Furthermore, the techniques of SSL can be combined with more established tools in the ML and DL toolbox. Another way a model such as NCEM might be augmented might be to enforce sparsity regularization on the cell signal sender-receiver matrix, to extract only the largest magnitude interactions or to impart the prior that cellular signaling is energetically expensive.

In the future, representations that are extracted using techniques in SSL may also be useful for enabling quantitative comparisons between cellular neighborhoods. A recent method called SCimilarity[14] was trained on non-spatial scRNA-seq data using triplet loss (mentioned above), with the explicit goal of optimizing a nonlinear latent variable projection such that Cell Ontology[15] relations were preserved. The authors show that the learned similarity function can be used to then query for cells similar to a generic input cell. A generally useful application of a similar idea may be to train a network in a similar fashion (using SSL) for generic querying of cellular neighborhoods or spatial gene expression patterns.

## Attention as a general dependency modeling operation and its potential in multimodal data integration

The attention mechanism is a critical component of the deep learning toolbox and in particular underlies the transformer[16], a highly successful architecture for a variety of applications. Turner[17] and Phuong and Hutter[18] give a substantive introduction to transformers and attention, but we will give a very abridged description here.

Attention is an operation on a sequence of "tokens", which are matrix representations for elements of a sequence or set, for example one vector (token) per word in a sentence or cell in a cellular neighborhood. Note that the token-wise representation is in itself a departure from common representations; for example previous methods might have encoded an entire sentence into a fixed-length vector, and instead here there would be a set of vectors representing components of a sentence. Self-attention is the variant of attention which parameterizes a feature update for each token in parallel using its own features and features from other tokens. This occurs using a learned kernel or similarity function between elements of the set, in what are commonly referred to as "key" and "query" projection operations (usually a single linear projection operation). At each layer, each token receives a feature update as a weighted average of its own and the other tokens' features. This is done by comparing its key projection with the query projections of all of the tokens (including itself), which gives a weighting factor over the set of tokens. This weighting factor, referred to as an attention coefficient, is then used to compute the weighted average of the features. The role of these learned similarity functions then is to allow the network to compare tokens in a dynamic fashion across the layers and use these similarities to compute a context-specific feature update for a given token given the others.

Adding complexity to this scheme is that attention often refers to "multi-headed attention", meaning that instead of updating every feature for every token at once, each "head" updates a subset of the features such that 2-headed attention would update half the features per head, and 3-headed attention would update one-third of the features per head and so on. Each head is parameterized by its own similarity function (in other words, each head is associated with a different query and key projection) which importantly is based on all of the features, incorporating a notion of global-to-local feature importance. Another variant is cross-attention, where feature updates are computed in terms of another set or sequence, for example updating tokens representing a radiology report using tokens representing a radiology image. The transformer architecture is unique in that its primary operation is composed attention operations for modeling generic sequences or sets, initially in natural language processing. An important note is that in computing pairwise-similarities between tokens, attention can be seen to be learning a graph over a set of elements.

A particular advantage of the self-attention mechanism is in its ability to learn an arbitrary dependency structure over data (in contrast to, for example, a message passing neural network, where the adjacency matrix is fixed). This capacity is generally useful for modeling dependencies amongst fixed items in, for example, natural language, or in amino acid sequences, but is also an important capability for multimodal data integration. We anticipate that this capacity will be explored extensively over the next few years as researchers employ the flexibility of attention to model arbitrary dependencies between modalities, elements of biological sequences (such as proteins or DNA sequences) or sets (such as cells in a spatial neighborhood).

One example of the relative ease of employing such a functionality, albeit in the clinical domain, is IRENE[19], a transformer based model for multimodal learning on paired radiology images, clinical text, and structured metadata. The authors use a method called Perceiver[cite] to allow for projection of large radiology images into a small set of tokens to reduce computational burden, but most importantly self-attention facilitates arbitrary feature interactions between tokens representing images, text, and metadata. With the appropriate datasets and training objectives, these approaches are a powerful tool to facilitate multimodal modeling of biological data at scale.

The attention mechanism itself is also a useful substrate for injection of human knowledge and expertise. The clearest example of this is in AlphaFold2 (AF2). One of AF2's hallmarks is a high degree of communication between a multiple-sequence (MSA) alignment processing operation and a "pair representation" which could be thought to encode structural information using a directed graph over residues.

The first way this is performed is via updates of the MSA features using information from the pair representation. The attention coefficient for a given pair of residues (i, j) in the MSA is given by a sum of a term representing affinities (query-key dot products) computed using the MSA and a term computed from the (i, j) features in the pair representation. One interpretation of this interaction is that the evolutionary couplings in the MSA ought to be a function of the structural or biophysical interaction of a pair of residues.

Another way human knowledge is incorporated into the optimization process is in the pair representation update. As mentioned, the pair representation can be thought to encode structural information. Therefore, the attention mechanism in AF2 is restricted in a mechanism referred to as triangle attention. Specifically, the idea that adjacent residues in a protein structure ought to inform each other. This is implemented by only allowing pairs of residues that share a residue (i.e. such that a sliding window of three residues should be allowed to attend to each other).

In the future, we believe that one impactful application of attention will be in examining contributions of different modalities to predictive modeling. For example, if paired radiology clinical notes and images were used to predict a complex phenotype. A practitioner could then develop a classifier that learns latent features for radiology and text, using a sparse attention for update of image features using text features prior to prediction. By inspecting the attention weights, an empirical understanding could be developed regarding the importance of text features, as some examples may not require a significant update using text features for accurate prediction.

# Conclusions

In this perspective, we provide an overview of opportunities for increased usage of DL tools for biology discovery. DL is already a critical tool in a variety of subfields of computational biology, particularly in single cell biology and protein biophysics. As new and flexible tools are developed for more fundamental computer science applications, we expect they will find fruitful application in biology.

# References

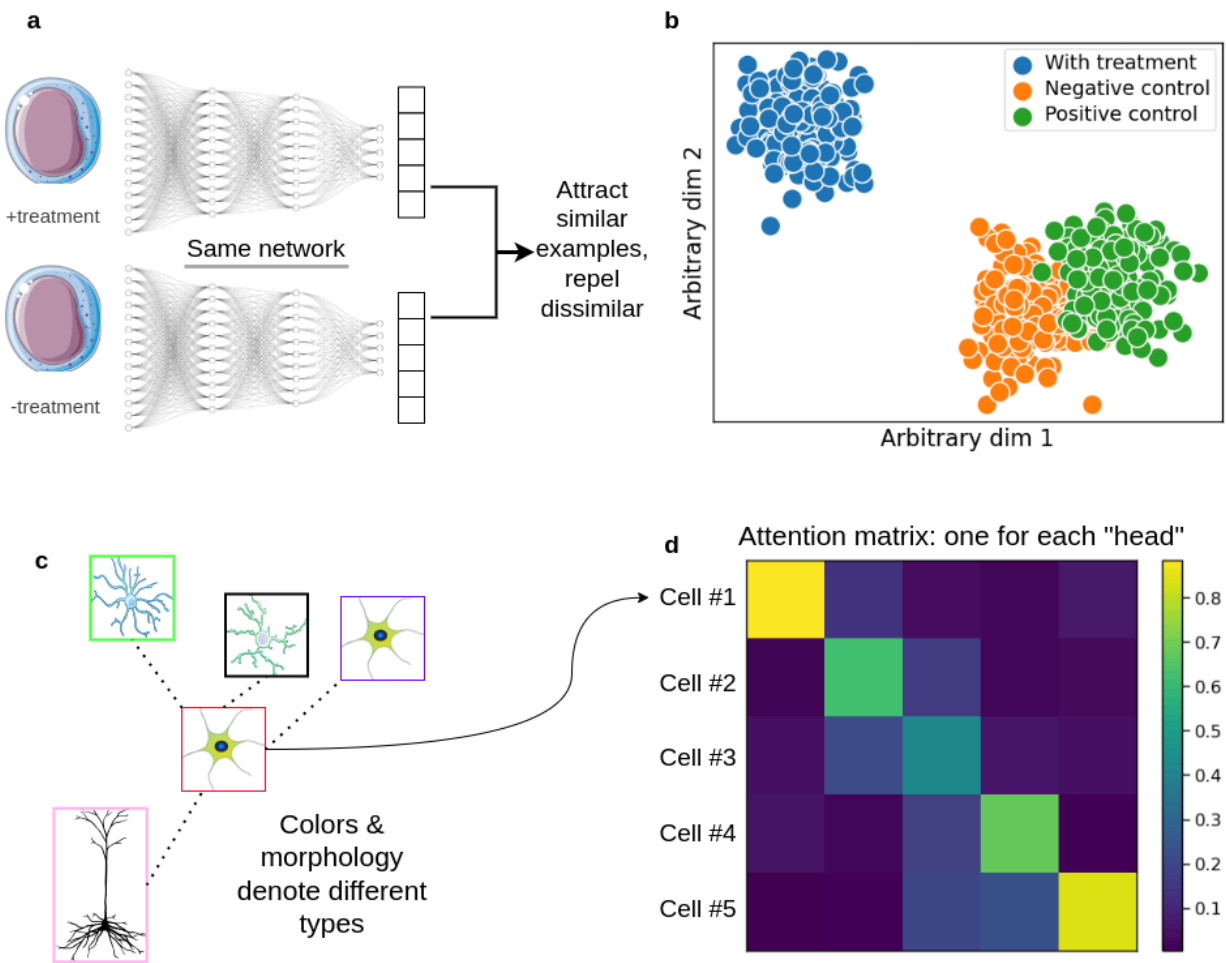

Figure 1. Simple cartoons of methods described in this perspective. A.Scheme of contrastive learning, wherein a neural network is used to encode different data points into some latent dimension. The network is optimized such that different data points (e.g. those that receive different drug treatments) are far away in latent space and those that are similar (e.g. those that receive the same drug treatment) are close together. B. A trivialized example of how a contrastive learning scheme might impose a useful geometry on an imagined dataset of treated cells and cells with positive or negative controls. C. A cartoon of the self-attention mechanism instantiated on a group of cells. The center cell, given in the red background, receives latent

features as a weighted average over the other cells. This pairwise similarity between elements (cells) is parameterized by a query and key projection of each cell's features and amounts to computing a directed graph over the cells as shown in d. D. A prospective attention matrix, where in practice there would be several such attention matrices. In this hypothetical case, the cells' features are not updated strongly by other cells, due to the strong diagonal. However, different cells have stronger off-diagonal elements than others.